\documentclass[aps,prl,twocolumn,showpacs,groupedaddress]{revtex4}
\usepackage {graphicx,epsfig,graphics,color,ams}
\draft

\begin{document}

\title{Observation of localized ferromagnetic resonance in a continuous
ferromagnetic film via magnetic resonance force microscopy}

\author{E. Nazaretski$^{1}$, D.V. Pelekhov$^{2}$, I. Martin$^{1}$, M. Zalalutdinov$^{3}$,\\
D. Ponarin$^{4}$, A. Smirnov$^{4}$, P. C. Hammel$^{2}$, and R. Movshovich$^{1}$}

\affiliation{$^1$Los Alamos National Laboratory, Los Alamos, NM
87545 USA,  $^2$Department of Physics, Ohio State University,
Columbus OH 43210,  $^3$ SFA Inc., Crofton, MD 21114,
$^4$ Department of Chemistry, North Carolina State University,
Raleigh, NC 27695}

\date{\today}

\begin{abstract}

We present Magnetic Resonance Force Microscopy (MRFM) measurements
of Ferromagnetic Resonance (FMR) in a 50 nm thick permalloy film,
tilted with respect to the direction of the external magnetic field.
At small probe-sample distances the MRFM spectrum breaks up into
multiple modes, which we identify as local ferromagnetic resonances
confined by the magnetic field of the MRFM tip. Micromagnetic
simulations support this identification of the modes and show they
are stabilized in the region where the dipolar tip field has a
component anti-parallel to the applied field.

\end{abstract}

\maketitle

Magnetic resonance force microscopy (MRFM) \cite{s:apl91} provides a
route to detection of magnetic resonance with excellent spin
sensitivity and spatial resolution and has received considerable
attention. Detection of electron spin resonance \cite{r:esr},
nuclear magnetic resonance \cite{r:nmr} and ferromagnetic resonance
(FMR) \cite{zhang:apl96, klein:loubens.prl2007:127601, r:fmrimage,
nazaretski:APL2007} with this approach has been reported. The high
sensitivity of MRFM when combined with the imaging mechanism of
Magnetic Resonance Imaging (MRI) allows excellent spatial resolution
in non-interacting spin systems \cite{r:90-nmresolutionimaging}.
Studies of patterned films show FMR can be detected in individual
microstructures with high sensitivity \cite{zhang:apl96, r:fmrimage,
midzor:jap00, klein:loubens.prl2007:127601}. As a consequence of the
strong interactions between spins in a ferromagnet the excitations
are collective (magnetostatic) modes, so the assumptions on which
MRI is based are invalidated, however recent work shows MRFM is also
a promising technique for spatially resolved FMR: Obukhov \emph{et
al.,} at OSU demonstrated that the large field gradient of the
micromagnetic probe enables ferromagnetic resonance imaging of the
magnetostatic modes of a 2 $\mu$m diameter permalloy dot with $\sim
250\; n$m resolution \cite{h:oboukhov.PRL.FMRdots2008}.

Here we argue that the micromagnetic probe \emph{locally stabilizes
magnetostatic modes}  detected in FMR thus indicating the potential
for scanned probe imaging of magnetic properties of extended
ferromagnetic films. Due to the strong interaction between the spins
the resonance modes are collective excitations of spins in the
entire sample, and in the case of the microfabricated samples, the
FMR modes are defined by the geometry of the sample.

We study the dependence of the mode resonance field on the
orientation of the film relative  to the external magnetic field,
and on the probe-sample separation. We find that localized FMR modes
are spatially confined immediately below the MRFM probe magnet. This
occurs because off the axis of the micromagnetic probe its dipolar
field is negative and, so {\rm reduces} the magnitude of the total
magnetic field stabilizing a confined mode. To understand the nature
of these modes and their behavior, we performed micromagnetic
simulations based on a finite element solution of the
Landau-Lifshitz (LL) equation with damping. The simulations reveal
that the degree of localization can be controlled not only by the
sample-probe separation, but that it is also very sensitive to the
tilt angle between the film and the applied magnetic field. These
results demonstrate confinement of the magnetostatic mode by the
localized inhomogeneous field of the probe, a mechanism distinct
from geometrical confinement such as was studied in
Ref.~\onlinecite{h:oboukhov.PRL.FMRdots2008}.

\begin{figure}[h]
\includegraphics [angle=0,width=6cm]{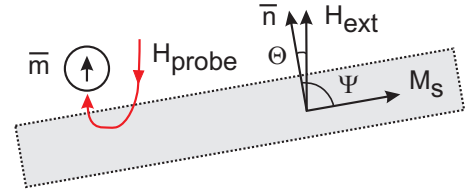}
\caption{(Color online) Schematic of the experiment. The sample is
tilted and placed in the uniform external magnetic field $\mathbf
H_{\rm ext}$. Here $\hat{n}$ is the normal to the sample surface,
$\Theta$ is the sample tilt angle with respect to $\mathbf H_{\rm
ext}$ and $\Psi$ is the tilt angle of the equilibrium magnetization
$\mathbf M_{s}$. The MRFM probe field is aligned with the direction
of $\mathbf H_{\rm ext}$ and its dipolar field reduces the magnitude
of $\mathbf H_{\rm ext}$ in the region off to the side of the
magnetic tip
 $\mathbf m$.} \label{fig:SetupScheme}
\end{figure}

The MRFM experiments were performed with the low temperature dual
MFM/MRFM apparatus at  Los Alamos National Laboratory. We used a
commercially available silicon nitride cantilever (resonance
frequency $f_c \approx 8.0$ kHz and spring constant $k \sim 10$
mN/m). The micromagnetic probe tip is a 2.4 $\mu$m diameter
spherical Nd$_2$Fe$_{14}$B particle. Its silicon nitride tip was
removed by focused ion milling, and the magnetic sphere was manually
glued to the cantilever in the presence of an aligning magnetic
field of a few kOe. The saturation magnetization of Nd$_2$Fe$_{14}$B
powder was determined to be $4 \pi M_s \approx 13.2$ kG at 10 K
using SQUID magnetometery. The spatial field profile produced by the
probe magnet has been carefully characterized
\cite{h:nazaretski.tipchar.apl2008}. For a detailed description of
the MRFM apparatus we refer the reader to
Ref.~\onlinecite{h:nazaretski.lt24}.

The 50 nm thick permalloy film was deposited on a 100 $\mu$m thick
silicon wafer on top  of a 20 nm thick Ti adhesion layer, and capped
with a protective 20 nm Ti layer. An approximately 1.9 $\times$ 1.9
mm$^2$ sample was glued to the strip line microwave resonator. The
temperature of the sample was stabilized at T = 11 K (temperature
stability was better than 5 mK) and the frequency of the microwave
field was set to $f_{RF}$ = 9.75 GHz. Fig.~\ref{fig:SetupScheme}
shows a schematic diagram of the experiment.

\begin{figure}[h]
\includegraphics [angle=0,width=8cm]{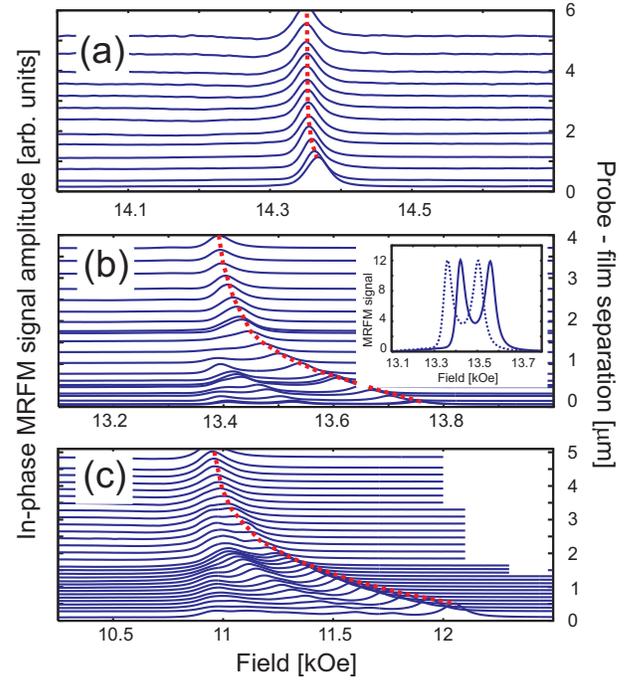}
\caption{(Color online) The in-phase component of the MRFM signal.
Each spectrum is plotted with an offset proportional to the
probe-sample separation. This separation can be read off of the
right-hand vertical axis far from resonance. Spectra were recorded
at three different angles $\Theta$: a) $1.0^\circ$, b) $3.4^\circ$
and c) $5.4^\circ$ at T = 11 K and $f_{\rm RF}$ = 9.75 GHz. Note the
different horizontal axis scales for each panel. The red dotted line
traces the evolution of the fundamental resonance mode. Spectra were
scaled for clarity of presentation. The inset in panel b) shows the
shift of the entire resonance structure with the change of the
microwave frequency from 9.75 to 9.65 GHz at a probe-film separation
of $\sim$ 760 nm.} \label{Figure 1}
\end{figure}

The in-phase component of the lock-in detected MRFM signal is shown
in Fig.~\ref{Figure 1} as a  function of the probe-sample separation
for three values of the tilt angle $\Theta$ between the normal to
the film plane and the direction of $\mathbf H_{\rm ext}$ (as shown
in Fig.~\ref{fig:SetupScheme}).  The sample tilt angle was
determined by measuring the deflection of a laser beam reflected off
the film surface at room temperature and found to be $1.0^\circ$,
$3.4^\circ$ and $5.4^\circ$.  While the absolute uncertainty of the
sample tilt at low temperatures is $\approx \pm1.0^\circ$, mainly
due to thermal contraction of the glue used to hold the sample in
place, the relative uncertainty of the tilt angles at low
temperature is better than 0.1 degree. These tilt angles were
confirmed by the measured values of the resonance field $\mathbf
H_{\rm ext}$ at large probe-sample separations (discussed further
below).

All spectra shown in Fig.~\ref{Figure 1} exhibit very similar
evolution as a  function of probe-sample separation. A single
fundamental resonance mode is observed at distances greater than a
few micrometers. The resonance field of this mode changes as the
probe approaches the sample surface (Fig.~\ref{Figure 1}a). At
larger values of $\Theta$ (Fig.~\ref{Figure 1}b, c) splitting of the
fundamental mode is observed at small separations. The magnitude of
the splitting increases with increasing $\Theta$ (Fig.~\ref{Figure
1}c).

\begin{figure}[h]
\includegraphics [angle=0,width=7cm]{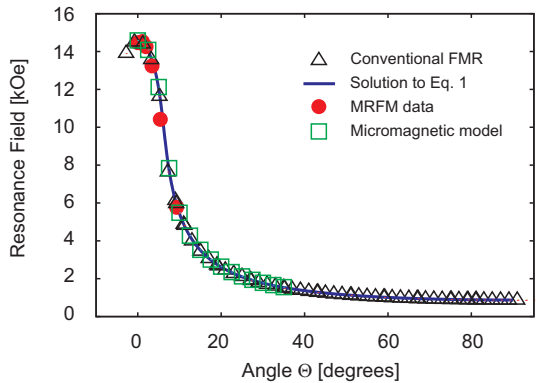}
\caption{(Color online) Angular dependence of the resonance field
for the uniform $\mathbf H_{\rm ext}$. Four data sets: conventional
FMR data ($\triangle$), solution to Eg. 1 (solid line), MRFM data at
large probe-sample separation ($\bullet$), and the results of the
micromagnetic modeling ($\square$) are in good agreement. The
numerical solution and the micromagnetic modeling were done using
the same simulation parameters: 4$\pi M_s$=11.3 kG, $f_{RF}$ = 9.5
GHz, $ \gamma = 2.89 \pm 0.05 $ GHz/kOe. MRFM data were acquired at
$f_{RF}$ = 9.75 GHz.} \label{AngularDependence}
\end{figure}

In order to better understand the evolution of the MRFM signal, we
performed FMR studies as a function of $\Theta$ in a conventional
FMR spectrometer. In Fig.~\ref{AngularDependence} we show the
angular dependence of the resonance field for the permalloy sample.
Conventional FMR experiments were performed at NCSU. The data were
obtained in a uniform applied field $\mathbf H_{\rm ext}$ and are
equivalent to the MRFM resonance spectra acquired at large values of
the probe-sample separation. In fact, the MRFM and conventional FMR
sets of data exhibit excellent agreement as shown in
Fig.~\ref{AngularDependence}. The value of the ferromagnetic
resonance field in a continuous film for a given $\Theta$ is
determined by the dispersion relation \cite{Wigen:ThinSolidFilms.1984}:

\begin{eqnarray}
  &\left( \frac{\omega}{\gamma}
  \right)^2=[H_{\rm ext}\cos(\Theta-\Psi)-4\pi M_s
  \cos(2\Psi)]\times\nonumber \\
  & [H_{\rm ext}\cos(\Theta-\Psi)-4\pi M_s \cos^2(\Psi)];\label{System1}\\
  &  H_{\rm ext}\sin(\Theta-\Psi)+2\pi M_s \sin(2\Theta)=0.\label{System2}
\end{eqnarray}
Here $\omega= 2 \pi f_{\rm RF}$ is the RF frequency, $\gamma$ is the
gyromagnetic  ratio, $4\pi M_s$ is the saturation magnetization and
$\Psi$  is the tilt angle of the equilibrium magnetization as shown
in Fig.~\ref{fig:SetupScheme}. We neglect the anisotropy
contribution because it is typically small ($\sim$ few Gauss) in
uniform permalloy films \cite{Frait:PermalloyFilms,
h:nazaretski.fmr.temperaturepermalloy.jap2007}. Equations
\ref{System1} and \ref{System2} cannot be solved analytically, so we
solved them numerically; the results are shown in
Fig.~\ref{AngularDependence} by the solid line. We used the
conventional FMR data set to verify the values of $4 \pi M_s = 11.3$
kG, and $ \gamma = 2.89 \pm 0.05 $ GHz/kOe. With conventional FMR we
never observed any multiple resonance structures reminiscent of
those appearing in Fig.~\ref{Figure 1}.

The origin of the isolated resonance modes seen in the MRFM studies
can  be understood in the following way. Conventional FMR theory
assumes a uniform magnetic field throughout the sample
\cite{FMR:Vonsovskii}. The LL analysis shows that the presence of a
position dependent inhomogeneous magnetic field of the probe may
result in a formation of localized FMR modes confined to the region
where the field inhomogeneity exceeds the value of $\Delta H_{max}
\simeq 2\pi M_s(L/a)$ (where $L$ is the thickness of the film and
$a$ is the radius of the localization)
\cite{h:oboukhov.PRL.FMRdots2008}. The MRFM probe produces an
approximately dipolar field in the plane of the sample. In the
simplest geometry ${\bf H}_{\rm ext}$ and the tip's magnetic moment
$\mathbf m$ are parallel to each other.  In this case the total
magnetic field directly under the tip (uniform external field plus
the dipolar field due to the tip) is higher than the external field,
but off to the side it is lower than ${\bf H}_{\rm ext}$ (see
Fig.~\ref{fig:SetupScheme}). In these two loci the field at which
resonance occurs is shifted to higher and lower fields, respectively
relative to resonance in the external field alone. While the modes
shifted down in field may overlap with the bulk magnetostatic modes,
the modes shifted up in field are well isolated, so they can readily
be used to perform local MRFM spectroscopy.

In order to quantify this picture we have conducted
micromagnetic simulations based on the Landau-Lifshitz (LL) equation
with damping:
\begin{equation}
\frac{\partial {\bf M}}{\partial t} = \gamma \mathbf M \times
\mathbf H_{\rm tot} - \frac{\lambda}{M_s^2}\mathbf M \times \mathbf
M \times \mathbf H_{\rm tot}, \label{LL}
\end{equation}
where $\lambda$ is the damping coefficient. The total magnetic field
$\mathbf H_{\rm tot}$ experienced by the sample magnetization
$\mathbf M$ consists of the external, probe, dipolar, exchange and
anisotropy fields. The numerical solution of Eq.~\ref{LL} for this
field configuration provides the information about the spatial
profile of the FMR modes excited in the sample and their
corresponding resonance fields. The details of the numerical
approach and its application to a variety of systems will be
presented elsewhere \cite{Pelekhov:Numerics}.

\begin{figure}[h]
\includegraphics [angle=0,width=7cm]{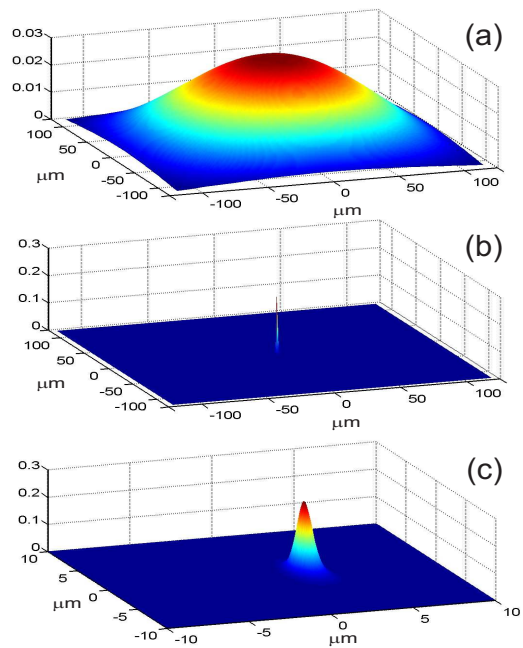}
\caption{(Color online) Numerically calculated spatial profile of
the excited transverse magnetization of the fundamental FMR mode in
a $250 \times 250 \; \mu m^2$  permalloy  film. The thickness of the
film is 50 nm and it is tilted at an angle $\Theta$ with respect to
the direction of the applied magnetic field. The probe magnet is a
2.4 $\mu$m diameter sphere placed at a distance $d$ over the center
of the film, (0,0,d) a) $\Theta$ = $0^\circ$, no probe magnet b)
$\Theta$ = $5.99^\circ$, $d$ = 1.0 $\mu$m, localized FMR mode at the
location of the probe can be seen. c) Zoom-in onto the localized
mode. Due to the sample tilt, the peak mode is localized to the side
of the probe. All modes were calculated for the value of $H_{\rm
ext}$ corresponding to $f_{RF}$ = 9.75 GHz.} \label{fig:ModeShapes}
\end{figure}

As numerical simulation of an infinite ferromagnetic film with  an
inhomogeneous external field profile is difficult we simulated a
finite ferromagnetic thin film that was large compared to the
phenomena of interest (250 $\times$ 250 $\mu m^2$ square with
thickness 50 nm). The magnetic parameters chosen for simulations
were obtained from the conventional FMR experiment: $4 \pi M_s$ =
11.3 kG and $\gamma$ = 2.89 GHz/kOe. For the damping we used
$\lambda$ = 0.005 $s^{-1}$, a typical value for permalloy.
Eq.~\ref{LL} was linearized on a variable density pixel grid with
higher pixel density in the vicinity of the probe magnet. The rare
earth probe magnet was modeled as a 2.4 $\mu$m diameter uniformly
magnetized sphere ($4 \pi M_s$ = 13.2 kG) placed at a height $d$
measured from the surface of the sphere. The probe was positioned
over the center of the film. Probe magnet parameters were
experimentally determined in Ref.~\cite{nazaretski:APL2007,
h:nazaretski.tipchar.apl2008}. The orientation of the probe magnetic
moment $\mathbf m$ is assumed to be aligned with the direction of
$\mathbf H_{\rm ext}$. The validity of the modeling approach was
verified by comparing the modeling results for a tilted film
(without MRFM tip) with the conventional FMR data (see
Fig.~\ref{AngularDependence}). In the absence of the localized probe
field, the fundamental (lowest frequency) resonance mode has a well
known bell-shaped spatial profile, as shown in
Fig.~\ref{fig:ModeShapes}a, and spans the entire sample. This result
is in good agreement with earlier theoretical work
\cite{kakazei:aplDotArrays}. In the presence of the probe magnet,
however, the mode changes its shape dramatically and is spatially
confined to the region immediately beneath the probe magnet (see
Fig.~\ref{fig:ModeShapes}b). A closer look at the mode localization
(see Fig.~\ref{fig:ModeShapes}c) reveals confinement on the order of
a few $\mu m^2$.  This is the region of the sample were the dipolar
field of the probe magnet opposes the externally applied field
$\mathbf H_{\rm ext}$. The asymmetry in the shape of the localized
FMR mode with respect to the position of the probe magnet is due to
the tilt of the sample, as illustrated in
Fig.~\ref{fig:SetupScheme}.

\begin{figure}[t]
\includegraphics [angle=0,width=0.8\columnwidth]{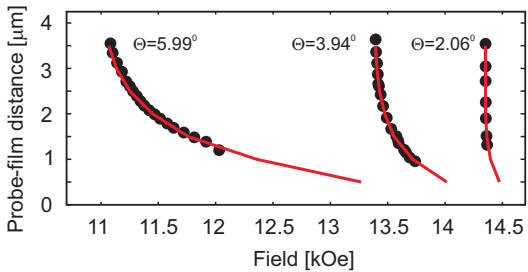}
\caption{(Color online) Comparison of the numerically calculated
evolution of the fundamental FMR mode (solid line) with data points
taken from Fig.~\ref{Figure 1}  (filled circles). The micromagnetic
modeling was performed for the tilt angles obtained from the
resonance fields of the fundamental mode at large probe-sample
separations. To obtain the agreement shown in
Fig.~\ref{fig:ExperimentModel} we had to increase the measured
probe-sample separation by $\approx$ 1 $\mu m$ (depending on the
tilt angle). The origin of the offset is the short range
probe-sample interaction which causes the cantilever to snap to the
film at probe-sample distances less than a micron. The nature of the
interaction has been discussed in
Ref.~\cite{Jean:vdwCapacitiveForceAFM.JAP99,
Hudlet:CapacitiveForceAFM.EuPJB.1998} and the magnitude of the
offset has been quantified in
Ref.~\cite{h:nazaretski.tipchar.apl2008}.}
\label{fig:ExperimentModel}
\end{figure}

In Fig.~\ref{fig:ExperimentModel} we compare the micromagnetic
simulations  of the evolution of the fundamental mode with the
experimental points taken from the dotted line in Fig.~\ref{Figure
1}. The values of the sample tilt angle $\Theta$ used in simulation
where chosen to match the values of the resonance field $\mathbf
H_{\rm ext}$ at large probe-sample separations. It can be seen that
the value of $\mathbf H_{\rm ext}$ shifts towards higher values as
the probe approaches the sample. This shift also increases with
increasing tilt angle of the sample. This is consistent with the
increase of the probe's negative field, which would require higher
values of $\mathbf H_{\rm ext}$ to satisfy the spin resonance
condition at $f_{RF}$ = 9.75 GHz.

In conclusion, we have observed FMR modes stabilized in a confined
region defined by the localized field of micromagnetic probe field.
These are observed when the sample is tilted with respect to the
orientation of the probe axis thus subjecting the sample to the
negative dipolar field produced off the axis of the probe magnet.
Our micromagnetic simulations accurately describe the experiments
and find the modes to be confined to a region of order a micron
across. Our experiments point to the possibility of spatially
resolved FMR (e.g. imaging and characterization of defects and
magnetic inhomogeneities) in continuous ferromagnetic films.

We thank Dr. T. Mewes for stimulating discussions and Dr. B. Houston
and Dr. J. W. Baldwin for sample fabrication. The work performed at
Los Alamos National Laboratory was supported by the US Department of
Energy, Center for Integrated Nanotechnologies. The work at OSU was
supported by the US Department of Energy through grant
DE-FG02-03ER46054. This work was also supported by the Office of
Naval Research through the Institute for Nanoscience at Naval
Research Laboratory. The work at NCSU was supported by NSF grant
ECS-0420775.

\end{document}